\documentclass[twocolumn,english,prb,amsmath,amssymb,preprintnumbers]{revtex4}
\usepackage[T1]{fontenc}
\usepackage[latin9]{inputenc}
\usepackage{amstext}
\usepackage{amsmath}
\usepackage{amsfonts}
\usepackage{amssymb}
\usepackage{mathtools}
\usepackage{graphicx}
\usepackage{rotating}
\usepackage {bm}
\usepackage[section]{placeins}
\usepackage{color}
\usepackage{multirow}
\usepackage{cellspace}
\usepackage{lipsum}
\makeatletter
\@ifundefined{textcolor}{}
{%
 \definecolor{BLACK}{gray}{0}
 \definecolor{WHITE}{gray}{1}
 \definecolor{RED}{rgb}{1,0,0}
 \definecolor{GREEN}{rgb}{0,1,0}
 \definecolor{BLUE}{rgb}{0,0,1}
 \definecolor{CYAN}{cmyk}{1,0,0,0}
 \definecolor{MAGENTA}{cmyk}{0,1,0,0}
 \definecolor{YELLOW}{cmyk}{0,0,1,0}
}

\def\kF{k_{\text{F}}}
\def\vF{v_{\text{F}}}

\def\NF{N_{\text{F}}}

\def\chis{\chi_{\text{s}}}

\def\epsilonF{\epsilon_{\text{F}}}

\def\vso{v_{\text{so}}}

\def\be{\begin{equation}}
\def\ee{\end{equation}}
\def\bea{\begin{eqnarray}}
\def\eea{\end{eqnarray}}
\def\bse{\begin{subequations}}
\def\ese{\end{subequations}}

\def\chis{\chi_{\text s}}

\def\vso{v_{\text so}}
\usepackage{dcolumn}
\usepackage{bm}
\usepackage{amsfonts}
\draft

\makeatother

\usepackage{babel}
\begin{document}
\preprint{arXiv:2601.19959}
\preprint{Accepted by PRB }

\bibliographystyle{unsrtnat}

\title{Comment on ``Instability of the ferromagnetic quantum critical point and symmetry of the ferromagnetic ground
        state in two-dimensional and three-dimensional electron gases with arbitrary spin-orbit splitting"}

\author{D. Belitz$^{1,2}$ and T. R. Kirkpatrick$^{3}$ }

\affiliation{
                 $^{1}$ Department of Physics and Institute for Fundamental Science, University of Oregon, Eugene, OR 97403, USA\\
                 $^{2}$ Materials Science Institute, University of Oregon, Eugene, OR 97403, USA\\
                 $^{3}$ Institute for Physical Science and Technology, University of Maryland, College Park, MD 20742, USA
                  }

\date{\today}
\begin{abstract}
Metallic quantum ferromagnets in the absence of quenched disorder are known to generically undergo a first-order 
quantum phase transition, avoiding the quantum critical point that had originally been expected. This is due to soft
modes in the underlying Fermi liquid that lead to long-ranged correlations. These correlations in turn yield a
nonanalytic dependence of the free energy on the magnetization even at a mean-field level that results in a 
fluctuation-induced first-order transition. Kirkpatrick and Belitz [Phys. Rev. Lett. {\bf 124}, 147201 (2020)]
have pointed out that one notable exception are non-centrosymmetric metals with a strong 
spin-orbit interaction. In such materials the spin-orbit interaction gives the relevant soft modes a mass, which
inhibits the mechanism leading to a first-order transition. Miserev, Loss, and Klinovaja [Phys. Rev. B {\bf 106}, 
134417 (2022)] have claimed that this conclusion does not hold if electron-electron interactions in the
particle-particle channel, or 2$\kF$ scattering processes, are considered. They concluded that this
interaction channel leads to soft modes that are not rendered massive by the spin-orbit interaction and again
lead to a first-order quantum phase transition. In this Comment we show that this conclusion is not correct
in three-dimensional magnets if the screening of the interaction is properly taken into account. 
\end{abstract}
%
%
\maketitle
Soft, or massless, modes and the long-ranged correlations they induce are important features of many-body
systems. A prominent example are soft particle-hole excitations in Fermi liquids.\cite{soft_mode_footnote}
Among other effects, they
are the origin of the nonanalytic temperature dependence of the specific heat,\cite{Carneiro_Pethick_1977}
as well as nonanalytic dependences of the spin susceptibility on the wave number\cite{Belitz_Kirkpatrick_Vojta_1997,
Chitov_Millis_2001, Galitski_Chubukov_Das_Sarma_2005} and on  the magnetic field.\cite{Misawa_1971,
Barnea_Edwards_1977, Betouras_Efremov_Chubukov_2005} It was shown in Ref.~\onlinecite{Belitz_Kirkpatrick_Vojta_1999}
that, as a result, the ferromagnetic quantum phase transition in clean metals is generically of first order. This
explains why a ferromagnetic quantum critical point, which originally had been studied as the epitome of
quantum phase transitions,\cite{Hertz_1976} is very hard to realize experimentally; for a review, see
Ref.~\onlinecite{Brando_et_al_2016a}. As a notable exception, it was pointed out by the current authors
that a spin-orbit interaction in non-centrosymmetric magnets gives the relevant soft modes a mass,
which renders inactive the mechanism leading to the first-order transition.\cite{Kirkpatrick_Belitz_2020}
This identified non-centrosymmetric metals with a sufficiently strong spin-orbit interaction as a promising
class for realizing a ferromagnetic quantum phase transition, and various candidate materials
were discussed. 

Miserev, Loss, and Klinovaja\cite{Miserev_Loss_Klinovaja_2022} have argued that 
(1) the mass-generating mechanism is not present in the particle-particle or 2$\kF$ scattering channel
(as is the case in disordered electron systems\cite{Altshuler_Aronov_1985, Belitz_Kirkpatrick_1994}),
(2) as a result, a ferromagnetic quantum critical point cannot be realized even in systems with a strong 
spin-orbit interaction, and (3) the sign of the leading nonanalyticity is not known, which casts further
doubt on efforts to identify the order of the transition. In this Comment we show that, while their 
claim (1) is correct, their conclusion (2) is not valid in three-dimensional ($3-d$) systems due to the 
Cooper screening of the relevant interaction amplitude.\cite{screening_footnote} As for claim (3), 
there are very general physical arguments for the sign of the nonanalyticity that all explicit calculations 
are in agreement with. 

To illustrate these points, we consider the single-particle Hamiltonian from Ref.~\onlinecite{Kirkpatrick_Belitz_2020},
\be
H_0 = \xi_{\bm k}\, \sigma_0 + \vso\, {\bm\sigma}\cdot{\bm k} - {\bm h}\cdot{\bm\sigma}
\label{eq:1}
\ee
Here $\vso$ is the spin-orbit coupling constant, and the momentum ${\bm k}$ in the spin-orbit term
can be replaced by a more complicated function that is odd in ${\bm k}$; this does not make a difference
for what follows. ${\bm\sigma} = (\sigma_1,\sigma_2,\sigma_3)$ are the Pauli matrices, and $\sigma_0$
is the 2$\times$2 unit matrix. ${\bm h} = (0,0,h)$ is an external magnetic field that we choose to point in
the $z$-direction, and $\xi_{\bm k} = \epsilon_{\bm k} - \mu$
with $\epsilon_{\bm k}$ the single-particle energy and $\mu$ the chemical potential. Diagonalizing this
Hamiltonian yields the Green function
\bse
\label{eqs:2}
\be
G_k = \frac{1}{2} \sum_{\beta=\pm} F_k^{\beta}\,M_{\beta}\left(\frac{\vso{\bm k}-{\bm h}}{\vert \vso{\bm k}-{\bm h}\vert}\right)
\label{eq:2a}
\ee
as a linear superposition of quasiparticle resonances
\bea
F_k^{\beta} &=& 1/\left(i\omega_n - \xi_{\bm k} - \beta \vert \vso{\bm k}-{\bm h}\vert\right)
\label{eq:2b}\\
                  &\approx& 1/\left(i\omega_n - \xi_{\bm k} -\beta\vso k + \beta{\hat{k}_z} h + O(h^2) \right)
\label{eq:2c}
\eea
where $k=(i\omega_n,{\bm k})$ combines a fermionic Matsubara frequency $i\omega_n$ with a
wave vector ${\bm k}$. The $M_{\beta}$ are 
\be
M_{\beta}(\hat{\bm e}) = (\sigma_0 +\beta {\bm\sigma}\cdot\hat{\bm e})\ ,
\label{eq:2d}
\ee
where $\hat e$ is an arbitrary unit vector. Equation~(\ref{eq:2c}) is valid for magnetic fields small compared to the
spin-orbit energy scale, $h \ll \vso \kF$, with $\kF$ the Fermi wave number. 
\ese
The relevant soft modes are convolutions of the quasiparticle resonances. In the particle-hole channel they are
of the form
\begin{widetext}
\bse
\label{eqs:3}
\be
\frac{1}{V}\sum_{\bm k} F_k^{\beta}\,F_{k-q}^{\beta'} \propto \frac{1}{i\Omega_m - \vF{\hat{\bm k}}\cdot{\bm q} + (\beta' - \beta)(\vso\kF - \hat{k}_z h)}
\label{eq:3a}
\ee
where $q = (i\Omega_m,{\bm q})$ with $\Omega_m$ a bosonic Matsubara frequency. This mode is soft (i.e., it diverges for 
small ${\bm q}$ and $\Omega_m$) if $\beta' = \beta$; however, a nonzero field does not cut off the singularity. In the terminology 
of Ref.~\onlinecite{Kirkpatrick_Belitz_2019a} this is a soft mode of the second kind with respect to the magnetic field, which 
cannot lead to a nonanalytic dependence of the free energy on the magnetization. For $\beta' = -\beta$ the spin-orbit interaction 
cuts off the singularity even in zero field, so the mode is massive and cannot lead to nonanalyticities either. This observation was 
the basis for the conclusion in Ref.~\onlinecite{Kirkpatrick_Belitz_2020} that a spin-orbit interaction can lead to a ferromagnetic 
quantum critical point. Now consider the analogous mode in the particle-particle channel, which is
\be
\frac{1}{V}\sum_{\bm k} F_k^{\beta}\,F_{-k+q}^{\beta'} \propto \frac{1}{2i\omega_n - i\Omega_m - \vF{\hat{\bm k}}\cdot{\bm q} 
                        + (\beta' - \beta)\vso\kF + (\beta' + \beta) \hat{k}_z h}\ .
\label{eq:3b}
\ee
\ese
\end{widetext}
This mode also is massive for $\beta' = -\beta$ and soft for $\beta'=\beta$, but here a magnetic field does cut off the singularity.
This is a soft mode of the first kind with respect to the magnetic field, which does lead to a nonanalytic dependence of the spin
susceptibility on the field, and of the free energy on the magnetization.\cite{Kirkpatrick_Belitz_2019a} 
This soft mode, which was not considered in Ref.~\onlinecite{Kirkpatrick_Belitz_2020}, was found to lead to a nonanalytic spin 
susceptibility of two-dimensional electron systems in Ref.~\onlinecite{Zak_Maslov_Loss_2010} and also considered again in
Ref.~\onlinecite{Miserev_Loss_Klinovaja_2022}. The existence of this soft mode of the first kind in clean metals is consistent
with the soft-mode structure of disordered electrons. There, it is known that a spin-orbit interaction renders massive
the spin-triplet soft modes (which are diffusive in this case), but leaves the spin-singlet ones soft, both in the particle-hole
and particle-particle channels. An additional magnetic field gives a mass to the spin-singlet modes in the particle-particle
channel, but not to those in the particle-hole channel, see, e.g., Table I in Ref.~\onlinecite{Belitz_Kirkpatrick_1994} 
This is the analog in disordered systems of what can be seen from Eqs.~(\ref{eqs:3}). 

\begin{figure}[t]
\includegraphics[width=7.5cm]{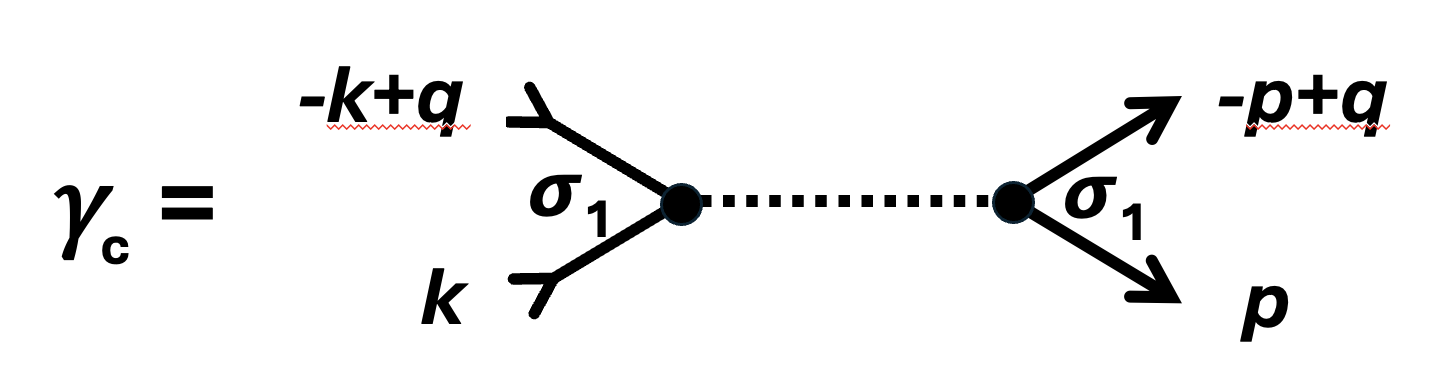}
\caption{Interaction amplitude $\gamma_c$ in the particle-particle spin-singlet channel, denoted by a dotted line. The 
vertices shown as filled circles carry a Pauli matrix $\sigma_1$.}
\label{fig:1}
\end{figure}

\begin{figure}[t]
\includegraphics[width=8.5cm]{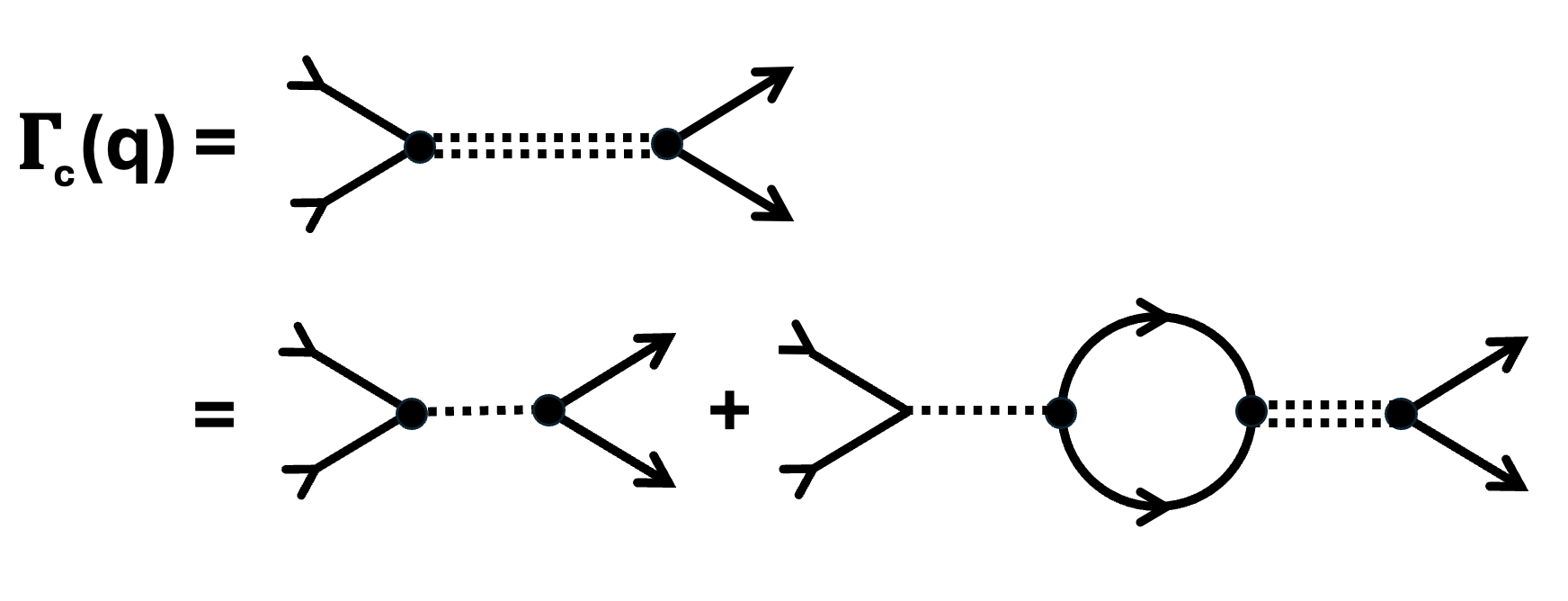}
\caption{The Cooper-screened particle-particle interaction amplitude $\Gamma_c(q)$,\cite{screening_footnote} denoted by a double dotted line.}
\label{fig:2}
\end{figure}

The soft modes discussed above are properties of the noninteracting electron system. In order for them to lead to 
nonanalyticities in observables an electron-electron interaction is required that provides the requisite frequency mixing.
The relevant bare interaction in the current context is the particle-particle channel or 2$\kF$-scattering interaction amplitude
shown in Fig.~\ref{fig:1}. For simplicity, we take this amplitude to be a constant $\gamma_c$. In order to avoid double
counting when the particle-hole interaction channels are also taken into account the hydrodynamic wave vector ${\bm q}$
must be restricted to small values $\vert{\bm q}\vert < \Lambda/\vF$ with $\Lambda$ an ultraviolet cutoff small compared to the
Fermi energy $\epsilonF$.\cite{Belitz_Kirkpatrick_2012a} This bare interaction amplitude
needs to be screened;\cite{Altshuler_Aronov_1985, screening_footnote} the relevant resummation is shown in Fig.~\ref{fig:2}. In the
absence of a spin-orbit interaction, $\vso=0$, and in zero field, we find analytically
\be 
\Gamma_c^{(0)}(q) = \frac{\gamma_c}{1 + 2\NF\gamma_c\left[\ln(\Lambda/\vF \vert{\bm q}\vert) + i_0(\Omega_m/\vF \vert{\bm q}\vert)\right]}
\label{eq:4}
\ee
Here $\NF$ and $\vF$ are the density of states at the Fermi level and the Fermi velocity, respectively, The function $i_0$ and two related functions of the scaling argument $\Omega_m/\vF\vert{\bm q}\vert$ are given by
\bse
\label{eqs:5}
\bea
i_0(x) &=& 1 - \arctan(1/x) - \frac{1}{2}\,\ln(1 + x^2)\ ,
\label{eq:5a}\\
i_2(x) &=& -1/(1+x^2)\ ,
\label{eq:5b}\\
i_4(x) &=& -(1-3x^2)/3(1+x^2)^3\ .
\label{eq:5c}
\eea
\ese
In a nonzero field we have, to $O(h^2)$,
\be
\Gamma_c(q) = \Gamma_c^{(0)}(q) \left[1 - 2 h^2 \frac{\NF\Gamma_c^{(0)}(q)}{(\vF\vert{\bm q}\vert)^2}\,i_2\left(\frac{\Omega_m}{\vF\vert{\bm q}\vert}\right) + O(h^4)\right]
\label{eq:6}
\ee

A crucial property of $\Gamma_c^{(0)}(q)$ is that it vanishes logarithmically as $\vert{\bm q}\vert \to 0$. This feature is
characteristic of the particle-particle channel and closely related to the structure of the BCS theory of superconductivity.
Technically, it arises from summing Eq.~(\ref{eq:3b}) over the fermionic Matsubara frequency. Crucially for what follows,
in the presence of a spin-orbit interaction there still is a contribution to $\Gamma_c(q)$ that is proportional to Eq.~(\ref{eq:4}),
although the prefactor involves spin traces that are more complicated than for $\vso = 0$. 

\begin{figure}[t]
\includegraphics[width=8.5cm]{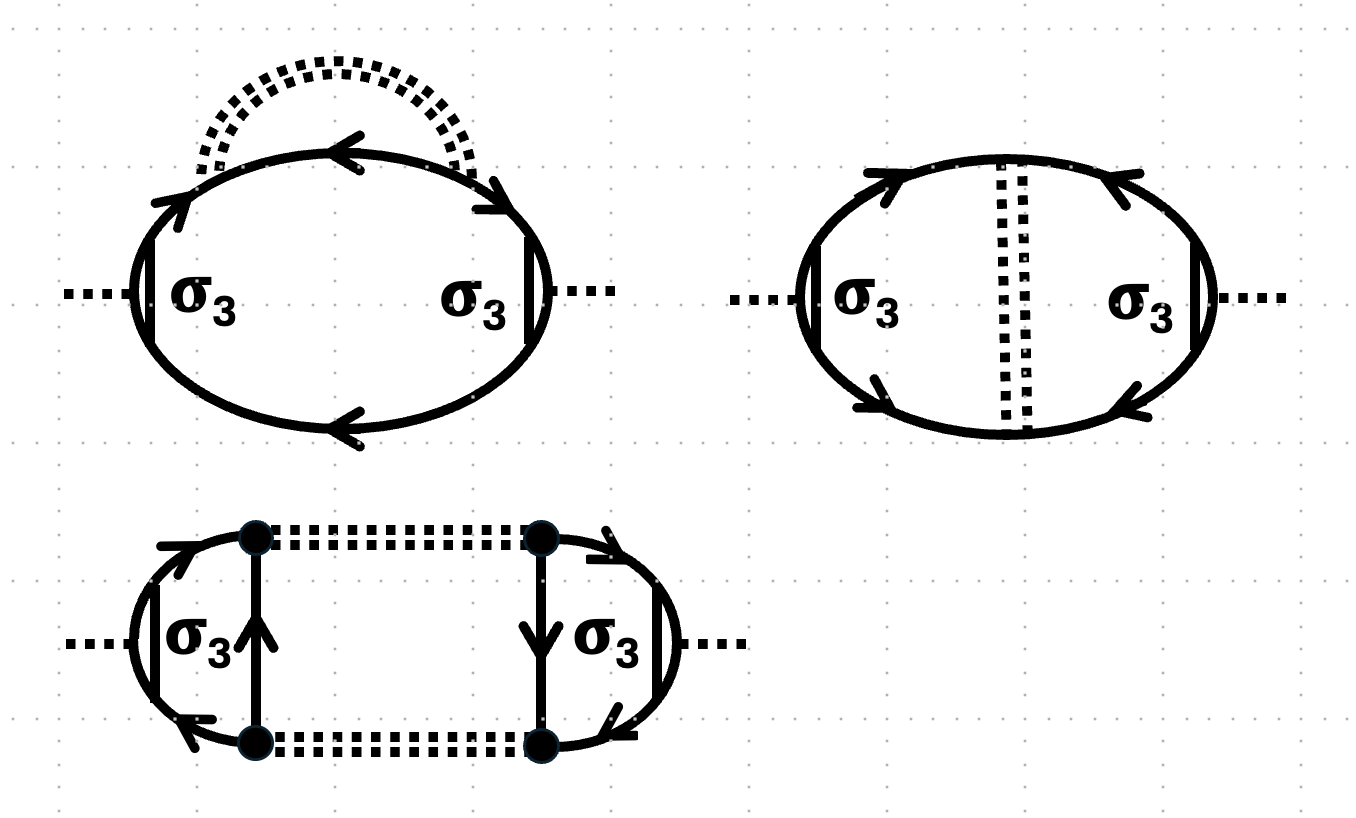}
\caption{One-loop contributions in the particle-particle channel to the spin susceptibility. The external vertex
              denoted by an open triangle carries a Pauli matrix $\sigma_3$.}
\label{fig:3}
\end{figure}

We now consider the contribution $\delta\chis^{\text{p}-\text{p}}$ of the soft mode in Eq.~(\ref{eq:3b}) to the spin susceptibility. 
As we will show below, it
suffices to perform the calculation for the case without a spin-orbit interaction. We consider contributions that carry one
integral over the hydrodynamic frequency-momentum variable $q$. The corresponding diagrams are shown in Fig.~\ref{fig:3};
in a field theory\cite{Belitz_Kirkpatrick_2012a} they correspond to one-loop diagrams. 
Power counting shows that the leading field dependence of all diagrams is $h^{d-1}$ in generic dimensions,
with possible logarithmic terms multiplying the $h^2$ in $d=3$. Focusing on the latter case, it is most convenient to expand to $O(h^2)$ 
and then extract the leading singularity from the resulting integral. We find
\bea
\delta\chis^{\text{p}-\text{p}} &=& O(h^0) + 24 h^2 {\sum_q} \ \NF\Gamma_c^{(0)}(q) \, \frac{1}{(\vF\vert{\bm q}\vert)^4}
\nonumber\\
&& \hskip -40pt \times \left[-2 i_4(\Omega_m/\vF\vert{\bm q}\vert) + \NF\Gamma_c^{(0)}(q)
                      \left(i_2(\Omega_m/\vF\vert{\bm q}\vert)\right)^2\right]
\nonumber\\
&& + O(h^4)\ ,
\label{eq:7}
\eea
where $\sum_q = T\sum_{i\Omega_m} (1/V)\sum_{\bm q}$ and the wave-number part of the integral is to be interpreted as
restricted to $h < \vF\vert{\bm q}\vert < \Lambda$. The lower cutoff is a result of the expansion in powers of $h$, which leads 
to IR divergent integrals. Inspecting the integral we make several observations: (i) The contribution linear in $\Gamma_c^{(0)}$
vanishes by means of the frequency integration, since $\int_0^{\infty} dx\,i_4(x) = 0$. (ii) If $\Gamma_c^{(0)}$ were a constant,
the integral would be proportional to $h^2 \ln(\Lambda/h)$ with a positive prefactor, which is qualitatively the same as the 
particle-hole channel contribution.\cite{Belitz_Kirkpatrick_Vojta_1997, Betouras_Efremov_Chubukov_2005} (iii) The dependence
of the screened interaction amplitude on the wave number and frequency suppresses the particle-particle channel contribution
logarithmically, and the final result is
\bea
\delta\chis^{\text{p}-\text{p}} &=& 2\NF \left\{ O(h^0) + \left(\frac{h}{\epsilonF}\right)^2 \frac{1}{2\pi}  \left[ \frac{\NF\gamma_c}{1+\ln(\Lambda/\epsilonF)} \right.\right.
\nonumber\\
&& \hskip -30pt  \left. \left.- \frac{1}{\ln(\epsilonF/h)} + O\left(\frac{1}{\log^2(\epsilonF/h)}\right) \right] + O(h^4) \right\}
\label{eq:8}
\eea
The leading nonanalyticity is proportional to $h^2/\ln h$ with a prefactor that is independent of the cutoff $\Lambda$. It
is subleading to the analytic $h^2$ term and therefore does not lead to a first-order ferromagnetic quantum phase
transition. The fact that the nonanalyticity in the particle-particle channel is weaker than the one in the particle-hole channel
by a factor of $1/\ln^2 h$ was recognized already in Ref.~\onlinecite{Belitz_Kirkpatrick_Vojta_1997}, but the 
calculation given above, which explicitly demonstrates this, was not performed. 

The remaining question is how Eq.~(\ref{eq:8}) gets modified in the presence of a spin-orbit interaction. There are
only two possible mechanisms that could lead to a stronger nonanalyticity than in Eq.~(\ref{eq:8}). First, $\vso\neq 0$
might cut off the logarithmic singularity in the interaction amplitude, Eq,~(\ref{eq:4}), which would lead to a leading
nonanalyticity proportional to $h^2\ln(1/h)$. This is not the case precisely
because of the soft mode that survives the spin-orbit interaction and contributes to the fermion loop in Fig.~\ref{fig:2}.
An explicit calculation confirms this, as we already mentioned after Eq.~(\ref{eq:6}). Second, there might be a
contribution linear in the interaction amplitude $\Gamma_c(q)$ that does not vanish due to a frequency integration;
this would lead to a leading nonanalyticity proportional to $h^2 \ln\ln(1/h)$. Such a contribution does not exist;
the first two diagrams in Fig.~\ref{fig:3} are the only ones that contribute to linear order in $\Gamma_c$, with or
without a spin-orbit contribution. Consequently, $\delta\chis^{\text{p}-\text{p}}$ for $\vso \neq 0$ is still qualitatively
given by Eq.~(\ref{eq:8}), albeit with a different prefactor due to more complicated spin traces. 

The above considerations illustrate points (1) and (2) from the introduction, i.e., the existence of a soft mode in the 
particle-particle channel and its irrelevance or the ferromagnetic quantum phase transition in $d=3$. In $d=2$ power counting
shows that the leading nonanalyticity is $h/\ln h$, but determining the prefactor would require a more sophisticated analysis.
We finally turn to point (3) in the introduction, namely, the sign of the nonanalyticity. A basic physical argument for the
sign has been given many times; see, e.g., Refs.~\onlinecite{Belitz_Kirkpatrick_Vojta_1999} and \onlinecite{Brando_et_al_2016a},
and we repeat it here:
The nonanalyticity is the result of the fermionic quantum fluctuations that lead to the soft modes discussed above and weaken 
the tendency of the system to order ferromagnetically. Therefore, the leading fluctuation contribution to the magnetic susceptibility 
at zero field, wave number, and temperature is necessarily negative. A nonzero magnetic field (or an external wave number, or 
temperature) weakens the fluctuations by giving the soft modes a mass and therefore the leading field dependence of the 
fluctuation contribution is positive.This is consistent with all reported calculations, and also with the result given above:
the leading fluctuation contribution in the particle-particle channel is the analytic $h^2$ term, which is positive. The
leading nonanalyticity is a correction to this term, and therefore negative. For $d<3$ the situation is different: Here
the nonanalyticity is the leading fluctuation contribution, which is proportional to $h^{d-1}/\ln(1/h)$, and therefore
must have a positive prefactor. 

We emphasize that perturbation theory by itself can never establish the sign of the effect, and there have been 
occasional speculations in the literature to the effect that higher-order terms in perturbative calculations could flip the 
sign of the nonanalyticity and lead to a quantum critical point even in the
absence of spin-orbit effects. However, these are inconsistent with the very basic and general physical argument given
above and there is no evidence, either calculational or observational, that they are correct.

In summary, the authors of Ref.~\onlinecite{Miserev_Loss_Klinovaja_2022} are correct in pointing out that there
is a soft mode in the particle-particle channel that is not cut off by a spin-orbit interaction. However, contrary to
their claim the effects of this soft mode in $3-d$ systems are too weak to pre-empt the quantum critical point
discussed in Ref.~\onlinecite{Kirkpatrick_Belitz_2020}. This is due to the Cooper screening of the interaction amplitude
that was neglected in Ref.~\onlinecite{Miserev_Loss_Klinovaja_2022}. In $2-d$ systems, on the other hand, this soft mode
should result in interesting effects that warrant further investigation.  


\begin{thebibliography}{21}
\providecommand{\natexlab}[1]{#1}
\providecommand{\url}[1]{\texttt{#1}}
\expandafter\ifx\csname urlstyle\endcsname\relax
  \providecommand{\doi}[1]{doi: #1}\else
  \providecommand{\doi}{doi: \begingroup \urlstyle{rm}\Url}\fi

\bibitem[sof()]{soft_mode_footnote}
The origin of their masslessness in both noninteracting and interacting clean
  fermion systems has been discussed in detail in
  Ref.~\onlinecite{Belitz_Kirkpatrick_2012a}. It technically is the same as the
  origin of the corresponding diffusive modes in disordered fermion systems,
  see Refs.~\onlinecite{Schaefer_Wegner_1980, Belitz_Kirkpatrick_1997} and
  references therein.

\bibitem[Carneiro and Pethick(1977)]{Carneiro_Pethick_1977}
G.~M. Carneiro and C.~J. Pethick.
\newblock Finite-temperature contributions to the magnetic susceptibility of a
  normal fermi liquid.
\newblock \emph{Phys. Rev. B}, 16:\penalty0 1933, 1977.

\bibitem[Belitz et~al.(1997)Belitz, Kirkpatrick, and
  Vojta]{Belitz_Kirkpatrick_Vojta_1997}
D.~Belitz, T.~R. Kirkpatrick, and T.~Vojta.
\newblock Nonanalytic behavior of the spin susceptibility in clean fermi
  systems.
\newblock \emph{Phys. Rev. B}, 55:\penalty0 9452, 1997.

\bibitem[Chitov and Millis(2001)]{Chitov_Millis_2001}
G.~Y. Chitov and A.~J. Millis.
\newblock First temperature corrections to the fermi-liquid fixed point in two
  dimensions.
\newblock \emph{Phys. Rev. B}, 64:\penalty0 054414, 2001.

\bibitem[Galitski et~al.(2005)Galitski, Chubukov, and {Das
  Sarma}]{Galitski_Chubukov_Das_Sarma_2005}
V.~M. Galitski, A.~V. Chubukov, and S.~{Das Sarma}.
\newblock Temperature-dependent spin susceptibility in a two-dimensional metal.
\newblock \emph{Phys. Rev. B}, 71:\penalty0 201302, 2005.

\bibitem[Misawa(1971)]{Misawa_1971}
S.~Misawa.
\newblock Logarithmic field dependence of the susceptibility of a paramagnetic
  fermi liquid - the {Pd} problem.
\newblock \emph{Phys. Rev. Lett.}, 26:\penalty0 1632, 1971.

\bibitem[Barnea and Edwards(1977)]{Barnea_Edwards_1977}
G.~Barnea and D.~M. Edwards.
\newblock A theory of anomalies in the field dependence of the magnetization of
  strongly enhanced itinerant paramagnets.
\newblock \emph{J. Phys. F}, 7:\penalty0 1323, 1977.

\bibitem[Betouras et~al.(2005)Betouras, Efremov, and
  Chubukov]{Betouras_Efremov_Chubukov_2005}
J.~Betouras, D.~Efremov, and A.~Chubukov.
\newblock Thermodynamics of a fermi liquid in a magnetic field.
\newblock \emph{Phys. Rev. B}, 72:\penalty0 115112, 2005.

\bibitem[Belitz et~al.(1999)Belitz, Kirkpatrick, and
  Vojta]{Belitz_Kirkpatrick_Vojta_1999}
D.~Belitz, T.~R. Kirkpatrick, and T.~Vojta.
\newblock First-order transitions and multicritical points in weak itinerant
  ferromagnets.
\newblock \emph{Phys. Rev. Lett.}, 82:\penalty0 4707, 1999.

\bibitem[Hertz(1976)]{Hertz_1976}
J.~Hertz.
\newblock Quantum critical phenomena.
\newblock \emph{Phys. Rev. B}, 14:\penalty0 1165, 1976.

\bibitem[Brando et~al.(2016)Brando, Belitz, Grosche, and
  Kirkpatrick]{Brando_et_al_2016a}
M.~Brando, D.~Belitz, F.~M. Grosche, and T.~R. Kirkpatrick.
\newblock Metallic quantum ferromagnets.
\newblock \emph{Rev. Mod. Phys.}, 88:\penalty0 025006, 2016.

\bibitem[Kirkpatrick and Belitz(2020)]{Kirkpatrick_Belitz_2020}
T.~R. Kirkpatrick and D.~Belitz.
\newblock Ferromagnetic quantum critical point in non-centrosymmetric systems.
\newblock \emph{Phys. Rev. Lett.}, 124:\penalty0 147201, 2020.

\bibitem[Miserev et~al.(2022)Miserev, Loss, and
  Klinovaja]{Miserev_Loss_Klinovaja_2022}
D.~Miserev, D.~Loss, and J.~Klinovaja.
\newblock Instability of the ferromagnetic quantum critical point and symmetry
  of the ferromagnetic ground state in two-dimensional and three-dimensional
  electron gases with arbitrary spin-orbit splitting.
\newblock \emph{Phys. Rev. B}, 106:\penalty0 134417, 2022.

\bibitem[Altshuler and Aronov(1985)]{Altshuler_Aronov_1985}
B.~L. Altshuler and A.~G. Aronov.
\newblock Electron-electron interaction in disordered conductors.
\newblock In A.~L. Efros and M.~Pollak, editors, \emph{Electron-Electron
  Interactions in Disordered Systems}, page~1. North-Holland, Amsterdam, 1985.

\bibitem[Belitz and Kirkpatrick(1994)]{Belitz_Kirkpatrick_1994}
D.~Belitz and T.~R. Kirkpatrick.
\newblock The {A}nderson-{M}ott transition.
\newblock \emph{Rev. Mod. Phys.}, 66:\penalty0 261, 1994.

\bibitem[scr()]{screening_footnote}
By this we mean screening by particle-particle pairs as shown in
  Fig.~\ref{fig:2}. This is analogous to the Thomas-Fermi screening in the
  particle-hole channel that is facilitated by particle-hole pairs. Cooper
  screening has been considered in the context of disordered electrons, see
  Ref.~\onlinecite{Altshuler_Aronov_1985}, but the same mechanism is present in
  clean systems.

\bibitem[Kirkpatrick and Belitz(2019)]{Kirkpatrick_Belitz_2019a}
T.~R. Kirkpatrick and D.~Belitz.
\newblock Soft modes and nonanalyticities in a clean dirac metal.
\newblock \emph{Phys. Rev. B}, 99:\penalty0 085109, 2019.

\bibitem[\.{Z}ak et~al.(2010)\.{Z}ak, Maslov, and Loss]{Zak_Maslov_Loss_2010}
R.~A. \.{Z}ak, D.~L. Maslov, and D.~Loss.
\newblock Spin susceptibility of interacting two-dimensional electrons in the
  presence of spin-orbit coupling.
\newblock \emph{Phys. Rev. B}, 82:\penalty0 115415, 2010.

\bibitem[Belitz and Kirkpatrick(2012)]{Belitz_Kirkpatrick_2012a}
D.~Belitz and T.~R. Kirkpatrick.
\newblock Effective soft-mode theory of strongly interacting fermions.
\newblock \emph{Phys. Rev. B}, 85:\penalty0 125126, 2012.

\bibitem[Sch{\"a}fer and Wegner(1980)]{Schaefer_Wegner_1980}
L.~Sch{\"a}fer and F.~Wegner.
\newblock Disordered system with n orbitals per site: Lagrange formulation,
  hyperbolic symmetry, and goldstone modes.
\newblock \emph{Z. Phys. B}, 38:\penalty0 113, 1980.

\bibitem[Belitz and Kirkpatrick(1997)]{Belitz_Kirkpatrick_1997}
D.~Belitz and T.~R. Kirkpatrick.
\newblock Theory of many-fermion systems.
\newblock \emph{Phys. Rev. B}, 56:\penalty0 6513, 1997.

\end{thebibliography}


\end{document}